\documentclass[prl,twocolumn,showpacs,floatfix]{revtex4}
\usepackage{graphicx}

\def\t2#1{\underline{\underline{#1}}}
\newcommand{\be}{\begin{equation}}
\newcommand{\ee}{\end{equation}}
\begin{document}
\title{Strain versus stress in a model granular material: a Devil's staircase.}
\author{Ga\"el Combe and Jean-No\"el Roux}
\affiliation{Laboratoire Central des Ponts et Chauss\'ees,
58 boulevard Lef\`ebvre, 75732 Paris cedex 15, France
}
\begin{abstract}
The series of equilibrium states reached by disordered packings of rigid,
frictionless discs in two dimensions, 
under gradually varying stress, are
studied by numerical simulations. Statistical properties of trajectories in
configuration space are found to be independent of specific assumptions ruling granular
dynamics, and determined by geometry only. A monotonic increase
in some macroscopic loading parameter causes a discrete sequence of rearrangements.
For a biaxial compression, we show that, due to the statistical importance of such events of
large magnitudes, the dependence of the resulting strain
on stress direction is a L\'evy flight in the thermodynamic limit.
\end{abstract}
\maketitle
The mechanical properties of granular media are currently an active field of research,
both in the condensed matter physics and in the mechanics and engineering
communities~\cite{BJ97,WG97,HHL98}.

Granular packings close to mechanical equilibrium are traditionnally modelled,
in the framework of continuum
mechanics, with elastoplastic constitutive laws~\cite{Muirwood,HHL98}, \emph{i.e.,}
incremental stress-strain
relations. Such laws, despite their practical success, were never clearly related to
grain-level mechanics. Moreover, cohesionless
granular systems seem to be quite different from ordinary solids. Many experimental,
theoretical and numerical studies~\cite{WG97,CLMNW96,RJMR96,OR97b,MJN98}
have recently been devoted to the peculiar features of
stress transmission in granular systems at equilibrium, with correlations
over length scales significantly larger than the grain size.

Observations of displacement fields and strains,
as the system moves from one equilibrium to another, are scarcer. Systems of rigid grains
are expected to deform because of rearrangements of the packing, rather than
contact elasticity. How such rearrangements average to produce a macroscopic strain,
related to stress variations, remains rather mysterious.
The rather singular, unilateral form of the local interaction
in such systems led some authors~\cite{CWBC98} to expect quite unusual macroscopic
properties, for which the very concept of strain, so familiar in mechanics of solids, would be
irrelevant. 

Direct grain-level approaches are, in principle, possible by numerical simulations.
However, one has then to define a complete mechanical model to enable a
calculation of particle trajectories. In practice, dynamical
parameters ruling energy dissipation are often chosen according to computational
convenience as much as physical accuracy. It would be desirable to assess the influence of
such choices on the results.

The present numerical study addresses those problems, as follows.

\begin{figure}[htb!]
\vbox{
\begin{center}
\includegraphics[totalheight=4.2cm]{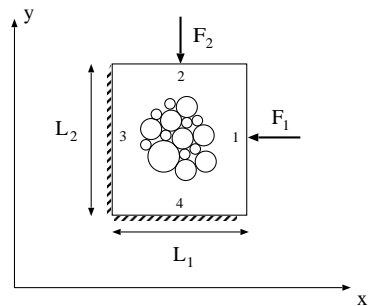}
\caption{Sketch of a biaxial compression test.}
\label{fig:biax}
\end{center}
}
\end{figure}

Disordered, dense assemblies
of rigid, circular, frictionless discs, are prepared by isotropic compaction. The force law
reduces to the condition that contacts transmit repulsive normal forces of unknown magnitude.
Then, the direction
of the load is gradually altered, thus simulating the biaxial test of fig.~\ref{fig:biax}.
Exploiting the isostaticity property~\cite{MO98a,TW99,JNR2000}
of such systems, which is exactly satisfied provided impenetrability is enforced accurately enough, 
we designed a prescription~\cite{JNR2000} for the computation of sequences
of equilibrium states, that we call the geometric
quasi-static method (GQSM), in which the only inputs are the geometric data. The strain versus
stress evolution in biaxial compression tests is recorded, and a statistical analysis of
fluctuations and system size dependence is carried out. Then the GQSM  predictions are compared
with those of a standard molecular dynamics (MD) method. 

First, samples of different sizes are generated (51 samples of N=1024 discs, 23 with N=1936,
10 with N = 3025 and 15 with N=4900).
Disc diameters are uniformly distributed between
0.5 and 1 (the largest diameter is chosen as unit length),
and packed in a loose state within a rigid, square box. After some amount of
random mixing (using some dynamical method with energy conservation) we proceed to the isotropic
compaction: two of the walls, 1 and 2 on fig.~\ref{fig:biax}, are now mobile, and submitted to 
compressive external forces $F_1$ (along the x axis on the figure) and $F_2 = F_1$ (along the y
axis). Stress components $\sigma_{11}=F_1/L_2$ and $\sigma_{22}=F_2/L_1$ are kept constant, equal
to $p$, while the system lengths along directions 1 and 2 ($L_1$ and $L_2$) decrease. 
To produce a dense, isostatic equilibrium state we use the `lubricated granular
dynamics' method of refs.~\cite{OR97b,JNR97b}. Then, the initial, reference configuration of the biaxial
experiment is ready: $F_2=(p+q) L_1^0$ is gradually increased while  $F_1=p L_2^0$ stays
constant, strain components are defined as the relative
decrease of lengths $(L_i)_{1\le i\le 2}$, with reference to their initial values $(L_i^0)_{1\le i\le 2}$
as $\epsilon_{ii} = -\Delta L_i / L_i^0$. One also defines the volumetric~\cite{3D}
strain as $\epsilon_v=-\epsilon_{11}-\epsilon_{22}$. We use units such that
$p=1$.

Our essential result here is the obtention of the $\epsilon(q)$ curves in the thermodynamic limit,
as loading parameter $q$ increases monotonically, at constant $p$. Let us first describe
the GQSM procedure. One 
starts from an equilibrium state in which the force-carrying structure is isostatic. This means
that the equilibrium conditions are sufficient to compute all contact force values, on the one hand
(the structure is devoid of hyperstaticity, it is not `overbraced' or `overconstrained'~\cite{MO98a}), and
that the force-carrying structure is rigid (devoid of mechanisms
or `floppy modes'~\cite{JNR2000}) on the other
hand. The first of these two properties stems from the condition that two grains need be exactly
in contact to transmit a force to one another~\cite{TW99,JNR2000}, and cannot interpenetrate.
It entails that force
values, once equilibrium positions are known, are geometrically determined, all material
properties being irrelevant in the limit of rigid grains. The second property is satisfied,
for stability reasons, because the grains are circular and contacts
do not withstand tension~\cite{JNR2000}.
It entails that an assembly of rigid discs will not deform at all
until some initially active contact opens. This cannot occur as long as contact forces are
compressive, since this would require the potential energy to increase from equilibrium. As soon
as one contact force vanishes, this contact will open~\cite{JNR2000},
because the resulting motion corresponds to an instability. Hence the following algorithm:

(1) At equilibrium, as $q$ increases from its initial
value $q_0$, the contact forces
depend linearly on $q$ (equilibrium equations are linear). When
$q$ reaches some value $q_0+\delta q$, the force vanishes in one contact, say $l_0$.

(2) Open $l_0$, all other contacts being maintained. Due to isostaticity, this
entirely determines the initial direction of motion for the whole structure. 
Keep moving the grains with the same prescription.

(3) When another contact, say $l_1$, closes, the new contact structure (the old one, minus
$l_0$, plus $l_1$) is isostatic and may carry the load with geometrically determined contact
forces. If there is no traction, a new equilibrium state has been found: go back to step 1.
Otherwise, pick up the largest traction, call the corresponding contact
$l_0$ and go back to step 2.

\begin{figure}[htb]
\vbox{
\begin{center}
\includegraphics[totalheight=4.8cm]{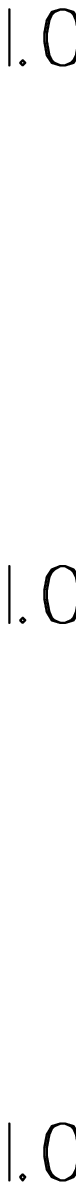}
\caption{`Axial' strain $\epsilon_{22}$ versus deviator $q$ in one sample with N=4900.}
\label{fig:esc}
\end{center}
}
\end{figure}

This procedure determines a series of equilibrium configurations, that are separated by
rearrangements occurring for discrete values of $q$. The
strain versus stress curve is a staircase (see fig.~\ref{fig:esc}).
As long as the same contacts carry the
load, the system does not deform; as soon as a rearranging event occurs, strain variables
jump to the values corresponding to the next equilibrium configuration. This algorithm clearly
involves, in steps 2 and 3, an arbitrary ingredient: the prescription that contacts open one by
one. The main merit of GQSM, however, is that it does not introduce parameters other
than geometric ones. 

We now focus on the rise of $\epsilon_{22}$ with $q$, close to the origin, and ask whether 
the staircase approaches a smooth curve in the thermodynamic limit. (Its initial slope, if finite,
would be the effective compliance of the material). To do so, one studies the statistics
of stress ($\delta q$) and strain ($\delta  \epsilon_{22}$, $\delta  \epsilon_{11}$) steps.

Successive $\delta q$ and  $\delta  \epsilon$ values are found to be independent, and the
width $\delta q$ of a stability interval is not correlated to the following
strain steps.
Throughout the investigated $q$ interval, 
the probability distributions of increments $\delta q$, $\delta  \epsilon_{22}$, and
$\delta W = p\delta \epsilon_v -q\delta \epsilon_{22}$, which is
the variation in potential energy corresponding to the current load, do not appreciably change~\cite{KS}.
No significant difference between samples is observed either.

\begin{figure}[htb]
\vbox{
\begin{center}
\includegraphics[totalheight=4.5cm]{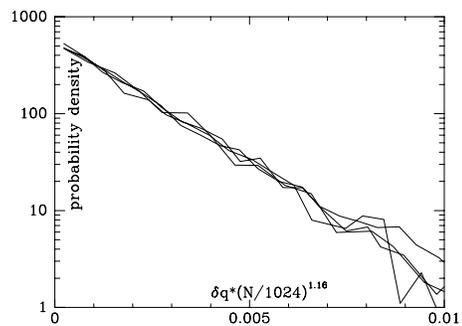}
\caption{Probability distribution function for rescaled stress increments $\delta q (N/1024)^{1.16}$
for the 4 values of system size $N$.}
\label{fig:dq}
\end{center}
}
\end{figure}

\begin{figure}[htb]
\vbox{
\begin{center}
\includegraphics[totalheight=5.3cm]{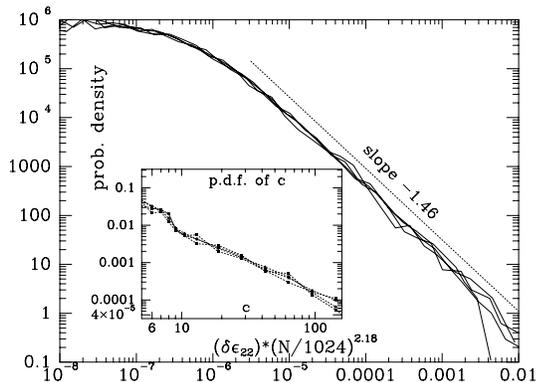}
\caption{Probability distribution function for rescaled strain
increments $\delta \epsilon_{22} (N/1024)^{2.18}$
for the 4 values of $N$.
The insert displays the probability distribution function of the number $c$ of lost contacts in one
strain step for each system size.}
\label{fig:deps}
\end{center}
}
\end{figure}

Both $q$ and $\epsilon_{22}$ values reached at a given stage can thus be regarded as sums of
equidistributed independent random increments. The distribution of stress increments $\delta q$
is displayed on fig.~\ref{fig:dq}.
It decays exponentially for large  $\delta q$, and is shifted
to smaller and smaller values as $N$ increases, so that the probability
distribution of the rescaled increment $\delta q N^\alpha$ is size-independent. We denote as
$\delta q_0$ its average. The exponent
can be estimated as $\alpha =1.16\pm 0.04$. Stability intervals 
shrink to zero as $N$ increases: in the thermodynamic limit,
any macroscopic load variation entails some motion of the grains.
This property of packings of
frictionless rigid grains, known as \emph{fragility}~\cite{CWBC98,JNR2000},
is unambiguously established by our simulations.
The value of $\alpha$ should be related to the
shape of the force distribution for small values, and to the varying sensitivity of the contact
forces to macroscopic load increments.
The classical central-limit theorem, applied to $q$ increments, will relate for large systems the
number of steps $M$ from the beginning of the biaxial compression to the current value of $q$, as
\be
M\sim {q\over \delta q_0} N^\alpha.
\label{eqn:clt}
\ee
The distribution of strain increments $\delta  \epsilon_{22}$ is shown on figure~\ref{fig:deps}.
A reduced variable $ N^\beta\delta \epsilon_{22} /\delta \epsilon_0$ can also be defined,
with an $N$-independent probability distribution. Remarkably, the (power-law distributed)
number $c$ of contact
losses (or gains) in one strain step, which tends to grow with $\delta \epsilon _{22}$, does not
show any significant size dependence, and contacts that simultaneously open or close
are homogeneously scattered throughout the sample, whatever $N$.
We estimated $\beta$ as
$\beta =2.18\pm 0.12$. Unlike for stress increments, the
distribution now decays as a power law:
$$p(\delta \epsilon _{22}) \sim (\delta \epsilon _{22})^{-(1+\mu)},$$
with $\mu = 0.46\pm 0.03$. As $\mu<1$, it should be remarked that, although the
\emph{typical} strain increment decreases as $N^{-\beta}$ in the limit of large samples, 
the \emph{average} strain increment does not exist. The standard central limit theorem
does not apply, but, due to the power law decay of the distribution function, one may
resort to a generalized central limit theorem~\cite{BG90}. In our case,
for large numbers $M$ of increments, the asymptotic form of their sum $\epsilon_{22}$ is
\begin{equation}
\epsilon_{22} \sim \delta \epsilon_0 N^{-\beta} M^{1\over \mu} \xi,
\label{eqn:Levy}
\end{equation}
with an $M$-independent random variable $\xi$ abiding by an asymmetric L\'evy distribution of index $\mu$.
 
\begin{figure}[htb]
\begin{center}
\includegraphics[totalheight=5.5cm]{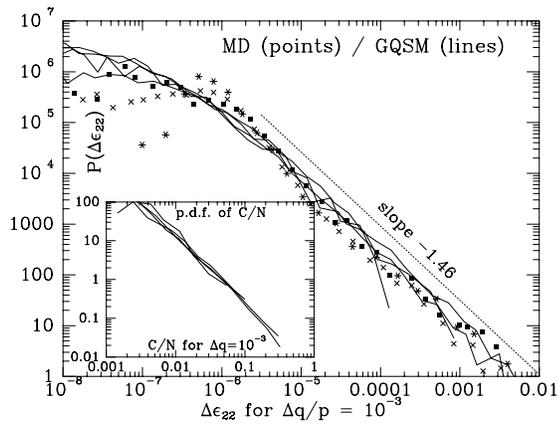}
\caption{Probability distribution function for strain increments $\Delta \epsilon_{22}$ corresponding to
$\Delta q = 10^{-3} p$ obtained for the 4 values of $N$ by GQSM (continuous curves)
and for N=1024 (crosses) and N=3025 (square dots) by MD. The insert shows the 
distribution of $C/N$, where $C$ is the
number of lost contacts in one increment $\Delta q$, for the different system sizes. 
}
\label{fig:Delta}
\end{center}
\end{figure}

The stress-strain relationship is obtained on combining eqns~\ref{eqn:clt} and~\ref{eqn:Levy}:
\begin{equation}
\epsilon_{22} = {\delta \epsilon_0 \over  (\delta q_0)^{1\over \mu}} q ^{1\over \mu}
N^{-\beta+{\alpha\over \mu}} \xi.
\label{eqn:const}
\end{equation}
The presence of $\xi$ in relation~\ref{eqn:const} implies that the strain versus stress curve
will never express a deterministic dependence. Some size effect -- a dependence on $N$ -- remains in the
thermodynamic limit, unless one has:
\begin{equation}
\alpha = \beta \mu
\label{eqn:abm}
\end{equation}
Our estimates of $\alpha$, $\beta$ and $\mu$ are compatible with this relation. If it
is satisfied, then the distribution of axial strain increments $\Delta \epsilon_{22}$
corresponding to a given, fixed, $q$ increment, 
$\Delta q$, should no longer depend on the system size as soon as $N$ is large enough for $\Delta q$ to
involve, typically, sufficiently many elementary rearranging steps. This was checked, confirming (see
fig.~\ref{fig:Delta}), within statistical errors, relation~\ref{eqn:abm} and the absence of size effects.

In order to compare the GQSM results to the predictions of more conventional methods,
we also simulated biaxial compressions on samples of 1024 and 3025 discs using MD, successively
imposing deviator increments $\Delta q = 10^{-3}$. Contacts obey an elastic unilateral law,
with a normal stiffness
equal to $10^5$. After each stress step, one waits for
a new equilibrium state, requesting forces on each grain to balance up to an accuracy of $10^{-5}$. 
Successive strain increments  $\Delta \epsilon_{22}^{MD}$ are uncorrelated and distributed according to
the same probability law, which coincides (within statistical uncertainties) as shown on
fig.~\ref{fig:Delta}, with that of increments  $\Delta \epsilon_{22}$ obtained for the same value of
$\Delta q$ by GQSM. Similarly, the(power-law) distribution of the
\emph{fraction} of the total number of contacts that change within one
fixed increment $\Delta q$ does not depend on $N$ (fig.~\ref{fig:Delta}).
We therefore conclude that, in a biaxial test,
the axial strain  $\epsilon_{22}$ dependence on deviator $q$,
as expressed by relations~\ref{eqn:const} and~\ref{eqn:abm},
is a L\'evy stochastic process.
It does not become deterministic in the thermodynamic limit, but it is devoid of size effect:
the strain versus
stress curve approaches a Devil's staircase with a dense set, on the $q$
axis, of discontinuities of random magnitudes.
Moreover, it appears not to depend on dynamical parameters (at least within
the accuracy of the present study) and to be
determined by the sole system geometry. These results apply,
without appreciable change in the statistics, throughout
the interval $0\le q/p \le 0.2$.

The evolution of $\epsilon _v$ (which can be of either sign, 
typically one order of magnitude smaller
than $\epsilon_{22}$)
is somewhat more
complicated, since successive increments $\delta \epsilon _v$
are not equidistributed (unlike $\delta W$).
Leaving a discussion of volumetric
strains to a future publication,
let us further comment here the behavior of axial~\cite{3D} strains.

Our results preclude the existence of a constitutive law in the usual sense. As $q$ is increased, 
$\epsilon _{22}$ is typically of order $q^{1/\mu}$, but its
actual value is essentially impredictible, with the remarkable
consequence that macroscopic models for granular mechanics should be of a stochastic,
rather than deterministic,
nature. Of course, such a behavior might be limited to frictionless grains,
and is thus perhaps more relevant
for some colloidal glasses than for sand, although important fluctuations and high noise levels
were sometimes reported for soil materials or glass beads. One does obtain an extremely noisy curve on
numerically compressing our system at constant strain rate instead of controlling the stress: 
understandably, between two equilibrium states that can be relatively far apart, the
contact structure does not consistently oppose any given level of stress,
and the observed response depends on
dynamical parameters. Laboratory tests, at constant strain rate, on such systems, 
would be sensitively influenced by the apparatus itself. It will be interesting
to study the dependence of parameter $\mu$ on grain shape and polydispersity.
Preliminary MD results on dense random packings of monodisperse spheres in 3D yielded
L\'evy-distributed large strain steps with a value of $\mu$ in the 0.4-0.6 range.

Physically, large strain increments are due to rearrangements involving many contact changes
(GQSM steps 2 and 3 have to be repeated). Had we resorted to the \emph{approximation
of small displacements} (ASD), in which all relevant quantities
are dealt with to leading order in the displacements, then,
as shown in ref.~\cite{JNR2000}, no iteration of steps 2 and 3 would have been necessary.
Within the ASD, contacts are replaced one by one, and we have checked that the resulting
strain increment distribution does possess an average value. This geometric approximation
would ensure~\cite{JNR2000} uniqueness of the equilibrium state and a one-to-one
correspondence between stress and strain. The ASD was used for slightly polydisperse
discs on a triangular lattice, in which case it can be justified
and constitutive laws are obtained~\cite{JNR2000,JNR97b} (the second ref.~\cite{TW99} 
proposes an independent implementation of the GQSM, with the ASD).
The distinctive mechanical features of model granular assemblies that are reported here
are thus due to the motion of the system in configuration space along a tortuous path between
local minima in a complex potential energy lanscape. Cooperative rearranging events
are also observed in glassy relaxation~\cite{DDKPHG98}, which might open interesting perspectives.

We are grateful to Jean-Philippe Bouchaud for useful suggestions.


\begin{thebibliography}{10}
\bibitem{3D} For convenience we resort to the vocabulary pertaining to
3D triaxial tests.
\bibitem{KS} Two variables are deemed equidistributed here
if the Kolmogorov-Smirnov likelihood estimator exceeds 0.05.
\end{thebibliography}
\end{document}